\documentclass[usenatbib,fleqn,twocolumn]{mn2e}
\usepackage{float}
\usepackage{mathptmx}
\usepackage{amsmath,amssymb,amsfonts,amsbsy} 
\usepackage[pdftex]{graphicx} 
\usepackage[backref=none]{hyperref}
\usepackage[usenames,dvipsnames]{color}
\usepackage{natbib}
 \usepackage[normalem]{ulem} 
\DeclareSymbolFont{cmletters}{OML}{cmm}{m}{it}
\DeclareMathSymbol{v}{\mathalpha}{cmletters}{"76}

\definecolor{MyDarkBlue}{rgb}{0,0.1,0.7}
\hypersetup{pdfborder={0 0 0},colorlinks,breaklinks=true,
  urlcolor={blue},citecolor={MyDarkBlue},linkcolor={magenta}}

\newcommand{\source}{GLEAM-X~J162759.5--523504.3}

\title[]{A precessing magnetar model for GLEAM-X J162759.5-523504.3}

\author[K.Y.~Ek{\c s}i \& S. {\c S}a{\c s}maz]{K.~Yavuz Ek\c{s}i$^1$ and Sinem {\c S}a{\c s}maz$^1$  \\
$^1$Istanbul Technical University,
  Faculty  of Science  and  Letters,  Physics Engineering  Department,
  34469,  Istanbul, Turkey, \\
  \href{mailto:eksi@itu.edu.tr}{eksi@itu.edu.tr}
}

\begin{document}

\maketitle

\begin{abstract}
We propose a precessing transient magnetar model for the recently discovered radio source GLEAM-X J162759.5-523504.3. We identify the observed period of $\sim 1$ ks as the precession period of the magnetar deformed due to its strong ($B_{\phi} \sim 10^{16}$~G) toroidal field. The resulting deformation of order $10^{-4}$ implies a spin period of $P_{\rm s} = 0.1$~s. Assuming a strong dipole field of $B_{\rm d} \sim 10^{14}$~G we predict a period derivative of $\dot{P}_{\rm s}\sim 10^{-11}$~s~s$^{-1}$. We also predict that the precession period of the magnetar can be observed in the hard X-ray band just as the other three galactic magnetars exhibit precession.
\end{abstract}

\begin{keywords}
stars: neutron -- stars: individual:GLEAM-X J162759.5-523504.3 -- radiation: radio
\end{keywords}

\section{Introduction}
\label{sec:intro}

The recent detection of a transient radio source with coherent emission in the radio 
band and high fractional polarization (88\%) clearly indicates to a neutron star magnetosphere as the origin \citep{hur+22}. 
The radio emission is modulated with a period of 18.18~min ($\simeq 1$~ks).
This is interpreted as the spin period of a rotationally-powered pulsar or a magnetar \citep{hur+22}. 

The spin period of rotationally-powered pulsars range between $P_{\rm s}\sim 10$~ms to $P_{\rm s}\sim 20$~s \citep[see][for the ATNF Pulsar Catalogue]{man+05}\footnote{See \url{https://www.atnf.csiro.au/research/pulsar/psrcat/}.}.
The spin period of galactic magnetars \citep[see][for a review]{kas17review} range between $P_{\rm s}=2-12$~s \citep[see][for the catalogue of galactic magnetars]{ola14}\footnote{See \url{http://www.physics.mcgill.ca/~pulsar/magnetar/main.html}.}.
The $\simeq 1$~ks period of the \source\ is unusual as the period of a pulsar or a magnetar with sufficient resources to exhibit the observed phenomena.

More recently, a fallback disc model was suggested for this source by \citet{ron+22}. Such discs were proposed to exist around anomalous X-ray pulsars (AXPs) \citep{cha+00} and all enigmatic young neutron stars \citep{alp01}. Such a disc was discovered around 4U~0142$+$61 by \citet{wan+06}.

Some of the galactic magnetars are observed to exhibit precession periods observed in the hard X-ray band:  4U~0142$+$61 with $P_{\rm p}=55$~ks \citep{mak+14,mak+19}, 1E~1547--54 with $P_{\rm p}=35$~ks \citep{mak+16,mak+21a} and SGR 1900$+$14 with $P_{\rm p}=40.5$~ks \citep{mak+21b}. \citet{mak21} notes that AXP~1RXS~J170849.0--400910 and SGR~0501$+$4516 likely exhibit the same behaviour. Such precession occurs owing to the deformation of the star by the strong toroidal magnetic fields \citep{tho95,dal+09,bra09}. Recently, precessing magnetars are proposed as central engines in short gamma-ray bursts \citep{suv21} and repeating fast radio bursts \citep{sob20,zan20,lev+20,was+21}.
In this \textit{letter} we interprete the 1.091~ks modulation of \source\ as the precession period of a young transient magnetar with a spin period of $P_{\rm s} \sim 0.1$~s. 
In the next section we introduce the model equations, and in the final section, we discuss the implications of this model and some predictions.

\section{Model equations}
\label{sec:model}

\begin{figure*}
\includegraphics[width=0.7\textwidth]{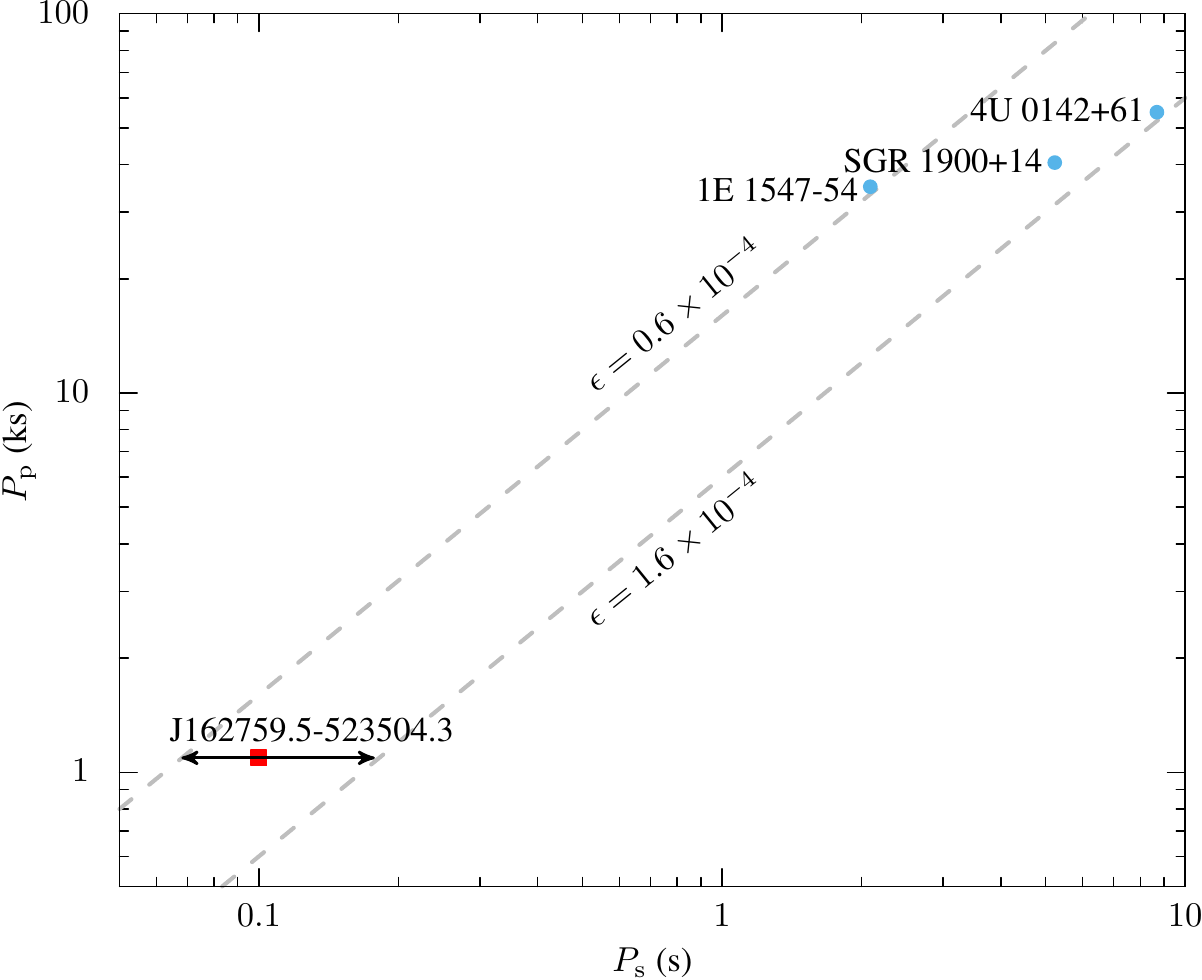}
\caption{The spin period $P_{\rm s}$ versus the precession period $P_{\rm p}$ of the three galactic magnetars as obtained by \citet{mak+14,mak+16,mak+21b} together with \source\ assuming 1.091~ks modulation is the precession period. The dashed grey lines correspond to asphericity of $0.6\times 10^{-4}$ compatible with 1E 1547--54 \citep{mak+16} and $1.6\times 10^{-4}$ compatible with 4U~0142$+$61 \citep{mak+14}.
\label{fig:pp}}
\end{figure*}

Assuming the object is spinning-down by magnetic dipole radiation torques its dipole field would be related to its period and period derivative as
\begin{equation}
B_{\rm d} = 6.4\times 10^{19} \sqrt{P_{\rm s}\dot{P}_{\rm s}}
\label{eq:B_p}
\end{equation}
\citep{gun69} and for a typical magnetar dipole field of $B_{\rm d} =10^{14}$~G one obtains the period derivative of the object as
\begin{equation}
\dot{P}_{\rm s} = 2.4\times 10^{-11} B_{\rm d,14}^2/P_{\rm s,-1}
\end{equation}
where we use the notation $X\equiv 10^n X_n$.
The spin-down power of the object is
\begin{equation}
L_{\rm sd} = 4\pi^2 I  \dot{P}_{\rm s}/P_{\rm s}^3
\end{equation}
where $I \simeq 10^{45}$~g~cm$^2$ is the moment of inertia. This can be expressed as 
\begin{equation}
L_{\rm sd} \simeq 10^{39} I_{45} B_{\rm d,14}^2/P_{\rm s,-1}^4~.
\label{eq:Lsd}
\end{equation}
This is compatible with the brightest radio pulses of the 
object being $L=4\times 10^{31}$~erg~s$^{-1}$ \citep{hur+22} since 
typically radio luminosity of pulsars are orders of magnitude smaller than their spin-down power.
Given the upper bounds on X-ray luminosity of the object $L_{\rm X}< 10^{32}$~erg~s$^{-1}$ as obtained by Swift \citep{hur+22} this renders the source as a \textit{rotationally powered magnetar}! This oxymoron is to mean that although the spin-down power is presently greater than the magnetic power of the object, when the source slows down sufficiently, its magnetic power will dominate over the spin-down power just as the typical galactic magnetars. The source is considered as a magnetar simply because of its strong toroidal field.

Magnetars are expected to have strong toroidal magnetic fields $B_{\phi} \sim 10^{16}$~G \citep{tho95,dal+09,bra09}.
Given the presence of magnetars with low dipole magnetic fields \citep{rea+10}, it is possibly this strong  toroidal field that makes a magnetar  different from a rotationally powered pulsar.
Magnetic stresses deform such a neutron star into a prolate spheroid 
\citep{iok01,cut02,iok04,has+08,mas+11,mas+13,mas+15,fri12,fre+21,zam21,sol+21}
with asphericity
\begin{equation}
\epsilon = \frac{I_1 - I_3}{I_3} \sim 10^{-4} B_{\phi,16}^2
\end{equation}
where $I_3$ is the moment of inertia around the magnetars symmetry
axis and $I_1$ is that around the axis orthogonal to this. 
As a result a magnetar will exhibit free precession with a period of
\begin{equation}
P_{\rm p} = P_{\rm s}/\epsilon
\end{equation}
\citep[see e.g.\ ][]{hey02}. In \autoref{fig:pp} we show the relation between the spin period and precession period of the three galactic magnetars together with \source. Assuming asphericity range of $\epsilon = (0.6-1.6) \times 10^{-4}$, as suggested by $P_{\rm s}/P_{\rm p}$ values of 1E~1547--54 \citep{mak+16} and of 4U~0142$+$61 \citep{mak+14}, we obtain a spin period in the range $P_{\rm p} \simeq (0.06-0.2)$~s.

\section{Discussion}
\label{sec:discuss}
We proposed a precessing magnetar model for the \source\ identifying the observed period of $\sim 1.1$~ks as the precession period. Assuming a typical toroidal magnetic field of $\sim 10^{16}$~G this implies a proloid deformation of order $\epsilon \sim 10^{-4}$ implying a spin period of $P_{\rm s} = 0.1$~s. This implies that the object is likely to be a magnetar on the basis of its strong toroidal field, but presently, due to its small spin period owing to its small age, its rotational power dominates over the X-ray luminosity. 

The upper bounds on the X-ray luminosity is 3 orders of magnitude smaller than the persistent X-ray luminosity of mature magnetars, $L_{\rm X} \sim 10^{35}$~erg~s$^{-1}$, but similar to quiescent state of transient magnetars. This implies that, assuming the above picture is correct, the source may show X-ray enhancement in the future.

Association of the object with galactic transient magnetars that show X-ray modulation due to precession \citep{mak+14}, we predict that the object might be brighter in hard X-rays compared to the soft X-ray band.

We predicted that the source should have a period derivative of $\dot{P}_{\rm s} \sim  10^{-11}$~s~s$^{-1}$ and spin-down power of $L_{\rm sd} \sim 10^{39}$~erg~s$^{-1}$. If the spin period of the source is 0.2~s, instead of the 0.1~s as assumed throughout the text, this luminosity drops 16 times owing to to the strong dependence on the period given in \autoref{eq:Lsd}.

\section*{Data availability}
No new data were analysed in support of this paper.

\section*{Acknowledgements}
We acknowledge support from T{\"U}B{\. I}TAK with grant number 118F028.

\footnotesize{
\bibliographystyle{mn2e}
\bibliography{refs} 
}

\end{document}